\documentclass[electronic]{vgtc}             

\ifpdf
  \pdfoutput=1\relax                   
  \usepackage{graphicx}                
  \DeclareGraphicsExtensions{.pdf,.png,.jpg,.jpeg} 
\else
  \usepackage{graphicx}                
  \DeclareGraphicsExtensions{.eps}     
\fi%

\graphicspath{{figures/}{pictures/}{images/}{./}} 

\usepackage{microtype}                 
\PassOptionsToPackage{warn}{textcomp}  
\usepackage{textcomp}                  
\usepackage{mathptmx}                  
\usepackage{times}                     
\usepackage{cite}                      
\usepackage{tabu}                      
\usepackage{booktabs}                  

\usepackage{mathtools}

\newcommand{\figref}[1]{Figure~\ref{#1}}

\newcommand{\tabref}[1]{Table~\ref{#1}}

\newcommand{\gemtwo}{Gemini$^2$}

\onlineid{0}

\vgtccategory{Research}

\vgtcinsertpkg

\title{Gemini$^2$: Generating Keyframe-Oriented Animated Transitions Between Statistical Graphics}

\author{Younghoon Kim\thanks{e-mail: yhkim01@uw.edu} \\
    \scriptsize{University of Washington} 
\and Jeffrey Heer\thanks{e-mail: jheer@uw.edu} \\ 
    \scriptsize{University of Washington}
}

\abstract{Complex animated transitions may be easier to understand when divided into separate, consecutive stages. However, effective staging requires careful attention to both animation semantics and timing parameters. We present \gemtwo{}, a system for creating staged animations from a sequence of chart keyframes. Given only a start state and an end state, \gemtwo{} can automatically recommend intermediate keyframes for designers to consider. The \gemtwo{} recommendation engine leverages Gemini, our prior work, and GraphScape to itemize the given complex change into semantic edit operations and to recombine operations into stages with a guided order for clearly conveying the semantics. To evaluate \gemtwo{}’s recommendations, we conducted a human-subject study in which participants ranked recommended animations from both \gemtwo{} and Gemini. We find that \gemtwo{}’s animation recommendation ranking is well aligned with subjects’ preferences, and \gemtwo{} can recommend favorable animations that Gemini cannot support.
} 

\CCScatlist{
  \CCScat{H.5.2}{User Interfaces}{User Interfaces}{Graphical user interfaces (GUI)}{};
  \CCScat{H.5.m}{Information Interfaces and Presentation}{Miscellaneous}{}{}
}

\begin{document}


\firstsection{Introduction}

\maketitle{}

Data analysis and communication often involve multiple statistical graphics and transitions among them~\cite{tableau_usage_pattern}. To convey the changes that occur in transitions, visualization researchers study and employ animation techniques that assess the effectiveness of animated transitions~\cite{tversky, anim_trend_vis} and their various strategies, such as staging~\cite{anim_transition}, staggering~\cite{staggering}, time distortions~\cite{siso}, and bundling (or path optimizing)~\cite{vector_field, trajectory}. Animation practitioners have created compelling media conveying data-driven insights, e.g., the famous Hans Rosling's presentation using Gapminder Trendalyzer~\cite{hans_rosling}, and an interactive article explaining machine learning algorithms~\cite{R2D3}.

However, authoring animated transitions between statistical graphics is not trivial; animation authors must implement low-level movements by  writing code that considers many design variations, such as staging and timing. Recently, researchers have facilitated this authoring experience using a non-programming interfaces~\cite{data_animator} or domain-specific languages~\cite{canis}. However, authors must still manually generate animation keyframes or specify changes by selecting low-level graphic components. 

In 2020, Kim \& Heer introduced Gemini~\cite{gemini}, a recommender system that suggests staged animations given start and end charts. However, Gemini's expressiveness is limited, as it cannot express independent intermediate states to the start and end charts. In addition, the Gemini's staged animation representations can be complex for users or systems to modify since they are combinations of the start and end chart specs with an animation spec in the Gemini grammar. 

\begin{figure}[!t]
  \centering 
  \vspace{-10pt}
 \includegraphics[width=\linewidth]{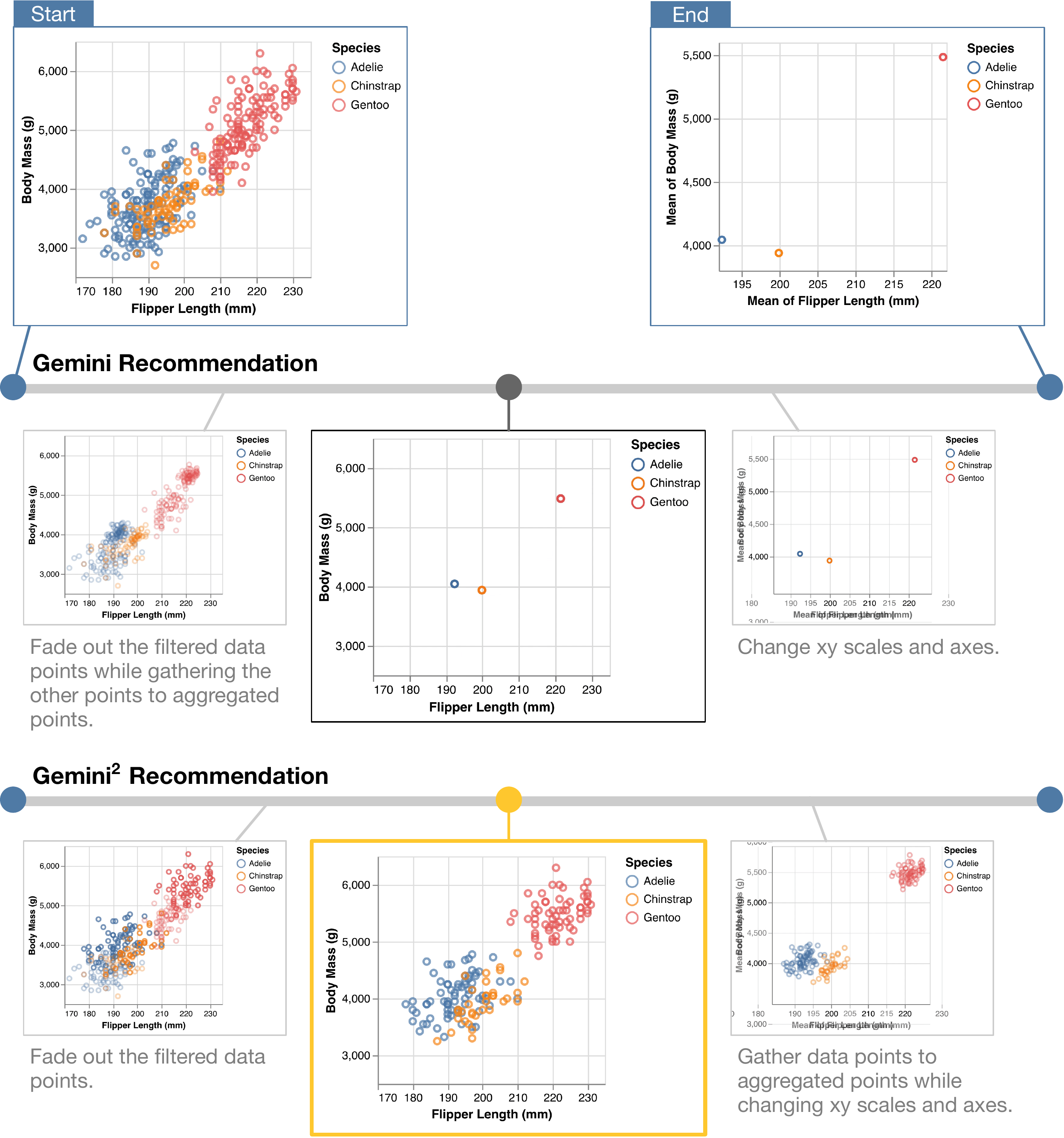}
 \vspace{-15pt}
 \caption{Staged animation timelines recommended by Gemini and \gemtwo{} for a transition where data are filtered and aggregated. The gray lines represents timelines, and the circles indicate start, end, and intermediate states. \gemtwo{} generates a keyframe chart (yellow circle) to separate two data transformations, which Gemini cannot do. }
 \vspace{-0.7cm}
 \label{fig:recom_comp}
\end{figure}

We introduce \gemtwo, a system for generating staged animated transitions between statistical graphics using keyframe sequences. Users first provide a chart array as a keyframe sequence. Then, \gemtwo{} automatically generates animations connecting consecutive keyframes. Users can then create nuanced animations by adjusting the animation specifications that connect each adjacent keyframe pair. \gemtwo{} thus supports a \textit{keyframe-oriented animation authoring approach}~\cite{anim_authoring_env}, where users design animations by importing or demonstrating keyframes identified in visualization specifications and connecting them with animation specifications. We describe \gemtwo{}'s enhanced expressiveness and ease-of-use vis-a-vis Gemini in Section 2.

Further, in Section 3 we demonstrate \gemtwo's utility by presenting a usage scenario that shows how it helps users create elaborate staged animations using keyframes. Additionally in this section, we evaluate keyframe recommendation through a crowd-sourced user study, where participants rank charts recommended by \gemtwo{} and Gemini. We find that while \gemtwo's recommendations are similarly compelling as Gemini's, \gemtwo{} can prioritize enumerated staged animations better than its predecessor. We conclude in Section 4 by discussing future directions, such as user interfaces, to improve the \gemtwo{} animation authoring process.

\section{Related Work}
\vspace{-3pt}
Animated transitions are commonly used to convey changes between two visualization states. Visualization researchers have shown that animations can facilitate a range of visual analysis tasks, from low-level value comparison and object tracking to higher-level tasks such as assessing trends and distributions~\cite{face2face, cone_tree, polyarchy, anim_mental_map, anim_decision_making, hops, hops_trend}. However, there is also justified skepticism of the effectiveness of animated transitions. Tversky et al.~\cite{tversky} found that some studies were biased in favor of animations, as the animation included additional information relative to other conditions. Further, they posed two high-level animation design principles, \emph{Congruence} (the structure and content of the external representation should correspond to the desired structure and content of the internal representation) and \emph{Apprehension} (the structure and content of the external representation should be readily and accurately perceived and comprehended). Robertson et al.~\cite{anim_trend_vis} found that though animated time-series data in multiple small charts was less effective in analysis tasks, subjects preferred animation in a presentation context.

Prior research has proposed and studied various animation techniques to enhance communication capability, including timing variations \cite{siso, staggering} and trajectory bundling~\cite{trajectory, vector_field}. To convey complex changes, animations adopt staging techniques that divide the given changes into multiple sub-sequences and animate them consecutively. Staged animations have outperformed unstaged variants for value estimation and change tracking~\cite{anim_transition}, conveying aggregate operations~\cite{agg_anim}, and understanding online dynamic networks~\cite{dynamic_network}. \gemtwo{} aims to aid the authoring of staged animations by generating them automatically from given keyframes and by recommending effective keyframes for given start and end visualization states.

Thompson et al.~\cite{anim_authoring_env} surveyed and characterize three animation authoring paradigms: \emph{keyframe animation} (setting keyframes and tweening them consecutively), \emph{procedural animation} (creating by behavioral parameters), and \emph{presets \& template animation} (applying pre-defined effects or templates). They found that most authors prefer keyframe animation, but animation authoring tools should support the other two paradigms depending on transition types and animation components. \gemtwo{} facilitates the keyframe animation paradigm by letting users  animate with customized keyframes and also by automatically suggesting keyframes.

A range of tools helps users create animated transitions, each making different trade-offs between \emph{ease-of-use} and \emph{expressiveness}. Focusing on ease-of-use, DataClip~\cite{dataclip} provides pre-defined animation templates, and gganimate~\cite{gganimate} lets users animate existing ggplot charts. Both limit expressiveness by using finite animation templates or making transitions within static visual encoding. In contrast, professional animation tools, such as D3 or Adobe After Effects, allow greater expressiveness but require significant effort, including writing program code to calculate and assign values for low-level components, diminishing ease of use. Between these extremes, researchers have developed declarative grammars for animated transitions, such as Canis~\cite{canis} and Gemini~\cite{gemini}. These grammars help users avoid imperative programming while focusing on specifying animation designs. However, Canis still requires users to make low-level selections using W3C Selector syntax~\cite{css} (like D3).

In contrast, Data Animator~\cite{data_animator} and CAST~\cite{CAST} enable keyframe animation authoring with a user-friendly graphical user interface (GUI). Users can set up keyframes by importing Data Illustrator~\cite{data_illustrator} projects or selecting graphic components. Then, users can configure animation designs using timing parameters and data mappings between adjacent keyframes. Likewise, \gemtwo{} lets users set keyframes by importing Vega-Lite visualizations. However, \gemtwo{} focuses on helping initiate the authoring process via keyframe recommendations, but does not provide a GUI for a general authoring experience.

\subsection{Comparison to Gemini}

\gemtwo{} enhances Gemini's expressiveness and ease-of-use. A limitation of Gemini is that its grammar cannot specify independent intermediate states to a start or end charts. For example, the intermediate states' data must be either the start or end chart's data.  In addition, the intermediate states' representations cannot be isolated; it requires users or systems to combine both start and end visualization specs with an animation spec in the Gemini grammar. \gemtwo{} instead permits intermediate state representations as independent visualization specifications, or keyframes. By doing so, \gemtwo{} increases expressiveness for staged animations and enacts the keyframe animation authoring paradigm.

For recommendations, \gemtwo{} incorporates GraphScape~\cite{graphscape} to suggest intermediate keyframes for a given pair of start and end charts. GraphScape is a formal model allowing machines to analyze transitions between statistical graphics, such as itemizing transitions into semantic edit operations, and measuring cognitive costs to following transitions. We extend GraphScape to itemize and recombine a given transition into multiple intermediate keyframes (charts), and prioritize each path (from the start to end through the intermediate charts) using heuristic rules. We enable these extensions by allowing GraphScape to synthesize a chart by applying edit operations. With GraphScape's assistance, \gemtwo{} can stage animations at a more fine-grained semantic level than Gemini can. For example, \gemtwo{} can divide a transition with a filter and aggregate operations into two stages by separating them, but Gemini cannot since it treats such data transforms as a single data change of a mark component (\figref{fig:recom_comp}).

\section{\gemtwo{}}

\gemtwo{} extends the Gemini grammar to support a keyframe-oriented animation authoring process. It represents an animation design as a sequence of $N$ keyframes (Vega-Lite~\cite{vega-lite} charts, denoted as $k_i$) and $N-1$ Gemini specifications($g_i$) and is formally written as
\[
  Animation := \{(k_1, k_2, ..., k_N), (g_{1\rightarrow{}2}, g_{2\rightarrow{}3}, ..., g_{N\rightarrow{}N-1})\}
\]
Keyframes are states that the animation passes through, and the Gemini specs are the specifications for each part of the animations connecting two adjacent keyframes. Therefore, users can express intermediate keyframes as arbitrary charts and connect them as an animation, which improves expressiveness relative to Gemini, as noted previously. Further, users can import multiple charts as keyframes and author staged animated transitions.

\gemtwo{} generates animation designs using the existing Gemini compiler; it creates $N-1$ animation objects for each consecutive chart pair $(k_i, k_{i+1})$ and corresponding Gemini spec ($g_{i\rightarrow{}i+1}$). The compiled $N-1$ animation objects are wrapped as a single object that plays the animations in a row.

\subsection{\gemtwo{} Recommendations}

\gemtwo{} provides two recommendation features by extending GraphScape and Gemini: (1) \textit{keyframe recommendations }for a given transition (start and end Vega-Lite charts), and (2) \textit{animation recommendations} for a given keyframe sequence. These two recommendation features let users automatically generate staged animations for a given transition or keyframe sequence.

\subsubsection{Keyframe Recommendations}

For given starting and ending Vega-Lite charts, \gemtwo{} can recommend intermediate keyframes to help authors explore staged animation designs. We implement these recommendations by leveraging GraphScape~\cite{graphscape}, a library for analyzing transitions between single-view Vega-Lite charts with Cartesian coordinates (e.g., bar, line, point charts). We extend GraphScape to generate charts; it now can synthesize charts by applying edit operations on existing charts. Also, we enable GraphScape to evaluate orders of edit operations in terms of how well they convey changes in an identifiable way.

\begin{figure}[!t]
  \centering 
  \includegraphics[width=\columnwidth]{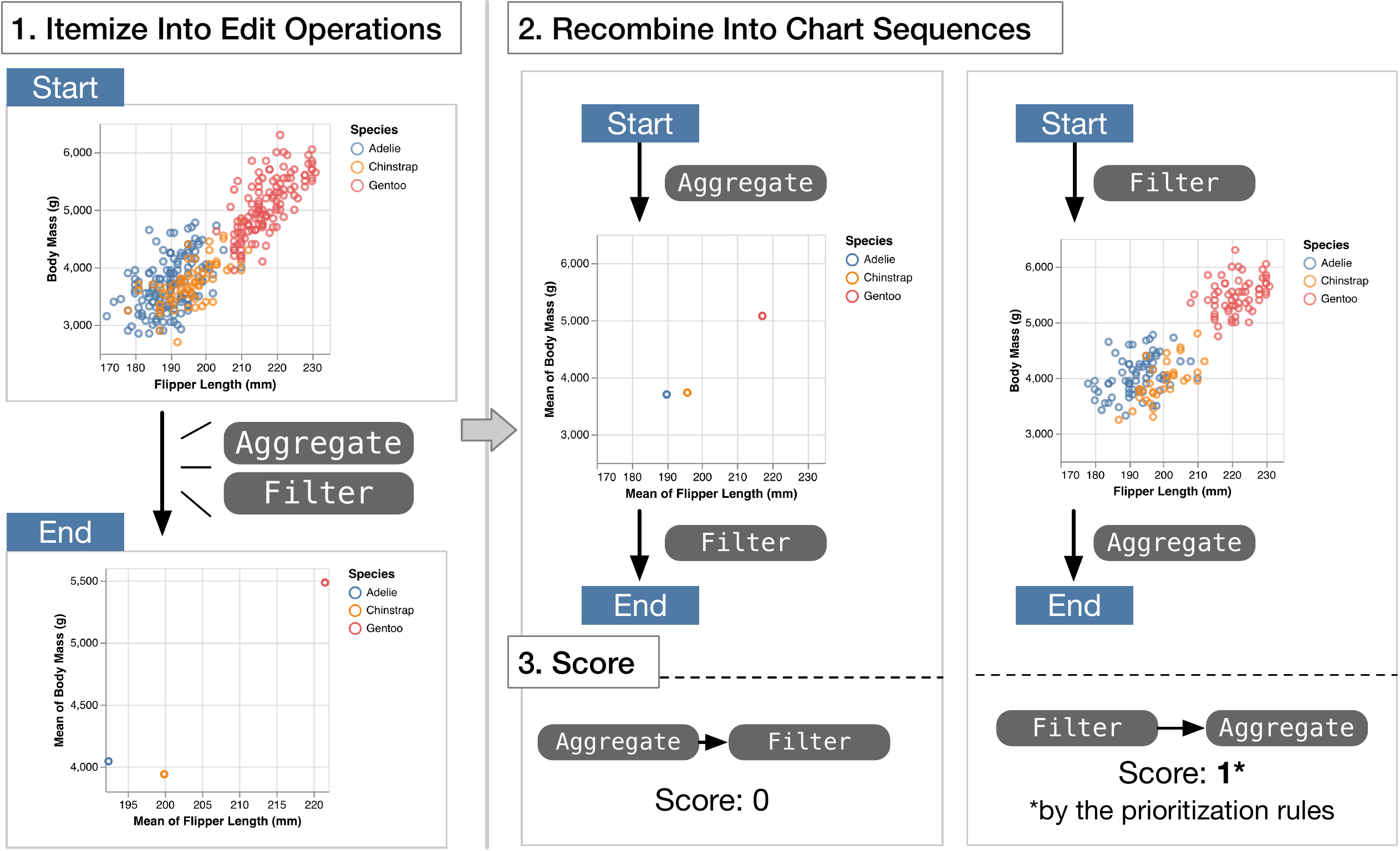}
  \vspace{-15pt}
  \caption{GraphScape's chart sequence recommendation process for a given transition. For given start and end charts, GraphScape enumerates  chart sequences by itemizing their changes into edit operations and recombining the operations. Then, it scores the sequences to rank them according to predefined prioritizing rules (see \tabref{tab:edit_ops_prioritization_rules}).}
  \label{fig:keyframe_recommendation}
  \vspace{-15pt}
 \end{figure}

GraphScape recommends keyframe sequences in three steps, as shown in \figref{fig:keyframe_recommendation}. It first itemizes changes from the start to the end into edit operations, which are atomic units that can be combined to represent any transition. Then, GraphScape enumerates intermediate charts by recombining and applying the itemized edit operations. During enumeration, GraphScape modifies the intermediate charts' scale domains (i.e., ranges of data values for corresponding visual properties) as the union of the given two charts to prevent unnecessary scale changes.

For example, imagine a transition filtering out some data and aggregating the remaining data into mean values. The start and end charts would have different scale domains ($domain_B \subset{} domain_A$). If an intermediate chart were applied only at the filter operation, the remaining data would have a different scale domain  ($domain_C \notin \{domain_A, domain_B\}$), and the chart sequence from A to C to B would change scale twice. However, by modifying $domain_C$ to  $domain_A$, the chart sequence changes scale only once.

After enumerating chart sequences, GraphScape prioritizes them based on how the edit operations are recombined. GraphScape applies a set of heuristic rules to evaluate the sequences. Each rule has (1) a \textit{condition} determining if certain edit operations should follow other operations or be placed together, and (2) a \textit{score} indicating if a sequence should be promoted or demoted if its operation order meets the condition. We define the rules in \tabref{tab:edit_ops_prioritization_rules}. Our goal is to make each edit operation identifiable by avoiding occlusion and implicit changes (such as filtering out some raw data when they are aggregated in charts). We hand-tuned the rule scores based on our experience and leave more rigorous tuning (e.g., data-driven adjustment in conjunction with user studies) as future work. We do not use GraphScape's transition costs in this prioritization process, another topic of potential future work. \gemtwo{} imports the GraphScape chart sequence recommendations as keyframes for staged animation.

\begin{table}[!t]
  \caption{Pre-defined rules for prioritizing edit operation order.}
  \label{tab:edit_ops_prioritization_rules}
  \scriptsize
  	\centering
  \begin{tabu}{ @{}p{8.3cm} }
  \toprule
   \textbf{Condition \& Description (Score)}\\
  \midrule
   \textbf{Filter $\rightarrow$ Aggregate/Bin} \newline
   Data should be filtered before being aggregated or binned as it is difficult to convey what points are being filtered out from an aggregate mark. Likewise, data should be disaggregated/unbinned before being filtered. (Score: 1)\\
  \midrule
   \textbf{Aggregate $\rightarrow$ Bin} \newline
   Data should \emph{not} be aggregated before being binned since data are aggregated by bin groups. Likewise, data should \emph{not} be unbinned before being disaggregated. (Score: -1)\\
  \midrule
   \textbf{Marktype $\rightarrow$ Aggregate} \newline
   Marktype should \emph{not} be changed before data are aggregated since some marktypes cannot support raw (un-aggregated) data, such as \texttt{bar} and \texttt{line} in Vega-Lite. Likewise, data should \emph{not} be disaggregated before changing marktypes. (Score: -1)\\
   \midrule
   \textbf{Add/Remove/Modify Encoding $\rightarrow$ Aggregate} \newline
   New encodings should be added before data are aggregated so that the aggregated distributions can be shown in the extended encoding. Likewise, data should be disaggregated before removing encodings. (Score: 1)\\
   \midrule
   \textbf{Modify Encoding with Scale} \newline
   When modifying an encoding channel, its corresponding scale changes should occur together so users can perceive these changes as a single field change. For example, changing an existing variable A on the x-axis to B while modifying its scale to log scale should occur synchronously so users can see  A $\rightarrow$ log(B) instead of A $\rightarrow$ B $\rightarrow$ log(B). (Score: 1)\\
   \midrule
   \textbf{Multiple Filters} \newline
   Multiple filter operations should \emph{not} take place together since viewers cannot separately identify each filter operation. (Score: -1)\\
   \midrule
   \textbf{Bin and Aggregate} \newline
   Bin should be conducted with aggregate operations. Otherwise, the binned data points cause occlusion. (Score: 1)\\
   \midrule
  \bottomrule
  \end{tabu}%
  \vspace{-20pt}
\end{table}


\subsubsection{Animation Recommendations with Keyframes}

\gemtwo{} can recommend animations for a given keyframe sequence ($k_i$) and number of stages ($M$). The formal representation is:
\begin{equation*}
  \begin{aligned}
  Anim.Recom: \{(k_1, k_2, ..., k_N), \, M\} \rightarrow{} (H_1, H_2, ...)
  \end{aligned}
\end{equation*}
where $H_i = (g^{i}_{1\rightarrow{}2}, g^{i}_{2\rightarrow{}3}, ..., g^{i}_{N-1\rightarrow{}N})$. The recommendation process leverages Gemini's recommendation feature. \gemtwo{} first gets staged animation designs for each consecutive chart pair through the Gemini recommendation feature
\[
  G_{i\rightarrow{}i+1} = (g^{1}_{i\rightarrow{}i+1}, g^{2}_{i\rightarrow{}i+1}, ...).
\]
In turn, it conducts a cross-join across each pair's recommendations. The cross-join enumerates arrays of Gemini specifications ($H_i$), each of which consists of $N-1$ specifications for every pair. \gemtwo{} includes only those where the total number of stages  equals the given number $M$.
\begin{equation*}
  \begin{aligned}
  Candidates = \{ H_i &| \forall i, \sum_{j}{stage(g^{i}_{j\rightarrow{}j+1})} = M \\
  & \& \; H_i \in G_{1\rightarrow{}2} \times G_{2\rightarrow{}3} \times ... \times G_{N-1\rightarrow{}N+1} \}
  \end{aligned}
\end{equation*}

The enumerated \gemtwo{} animation specifications are ranked by using Gemini's complexity measure:
\[f_{\text{total complexity}}(H_i) = \sum_{j=1} f_{\text{complexity}}(g^{i}_{j\rightarrow{}j+1}).\]
Gemini's complexity measure is a proxy for human effort to follow the animated changes. Gemini recommends staged animation designs with lower complexity measures. Similarly, \gemtwo{} uses the summation of the complexity as an evaluation metric. Doing so is still valid because (1) Gemini's complexity is designed to be summed across each stage, and (2) each consecutive chart pair forms a stage. As a result, \gemtwo{} prioritizes the Gemini specification array with the lowest total complexity score.

\subsection{Usage Scenario}

We now demonstrate how \gemtwo{} lets users create animations with keyframes through an example scenario. Here, a virtual author, K, creates  Kim et al.'s two versions of staged animations, \emph{staged elaborate} and \emph{staged basic}, conveying a median aggregate operation for a 1D data distribution~\cite{agg_anim}.

\begin{figure}[!t]
  \centering 
  \includegraphics[width=\columnwidth*1]{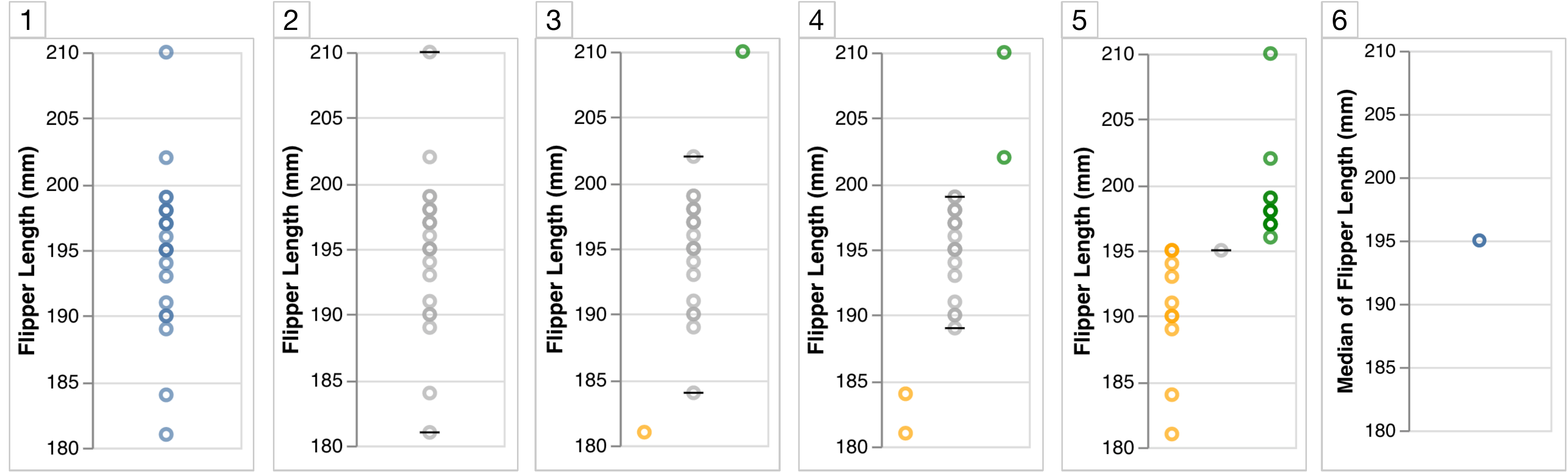}
  \vspace{-18pt}
  \caption{Example \gemtwo{} keyframes for the staged elaborate animation for the median aggregate operation~\cite{agg_anim}.}
  \label{fig:usage_scenario}
  \vspace{-18pt}
 \end{figure}

K starts to author \emph{staged elaborate} by demonstrating its keyframe sequence in Vega-Lite charts (\figref{fig:usage_scenario}). Once the keyframes are set up, K uses \gemtwo{} to automatically generate a basic staged animation through its recommendations for the keyframe sequence. Then, K finishes by editing a small amount of the generated animation specifications to control the timing of each stage, including pauses and durations, plus a staggering effect on 4$\rightarrow$5. The result can be simplified to a \emph{staged basic} by removing keyframe 2, 3, and 4.

This authoring process is not available in Gemini since it cannot support defining intermediate states with extra mark layers (2 to 6) that were not included in start and end charts.

\section{User Study}

We conducted a user study to assess whether \gemtwo{} recommended  more compelling staged animation designs than Gemini.

We recruited 50 participants (35 males, 15 females) from Amazon Mechanical Turk. In the study, we provided 4 different types of transitions between statistical graphics. For each transition, subjects ranked 5 corresponding animation designs: two were Gemini's best recommendations for 2- and 3-stage animations (\texttt{g-s2}, \texttt{g-s3}), and two were \gemtwo{}'s best recommendations for 2- and 3-stage animations (\texttt{g2-s2}, \texttt{g2-s3}) except for Stimulus 4. In Stimulus 4, \gemtwo{}'s best is the same as the Gemini's best; therefore, we used 2- and 3-stage animations with lower keyframe recommendation scores to check if the score also aligned with subjects' preference. The fifth design was an unstaged animation (\texttt{no-stage}) that directly interpolates all changes. We included the unstaged animation to determine if \gemtwo{} can recommend compelling staged animations for complex transitions better than the unstaged animations. More details are available on our supplementary material.

\begin{figure}[!t]
  \centering 
  \includegraphics[width=\columnwidth]{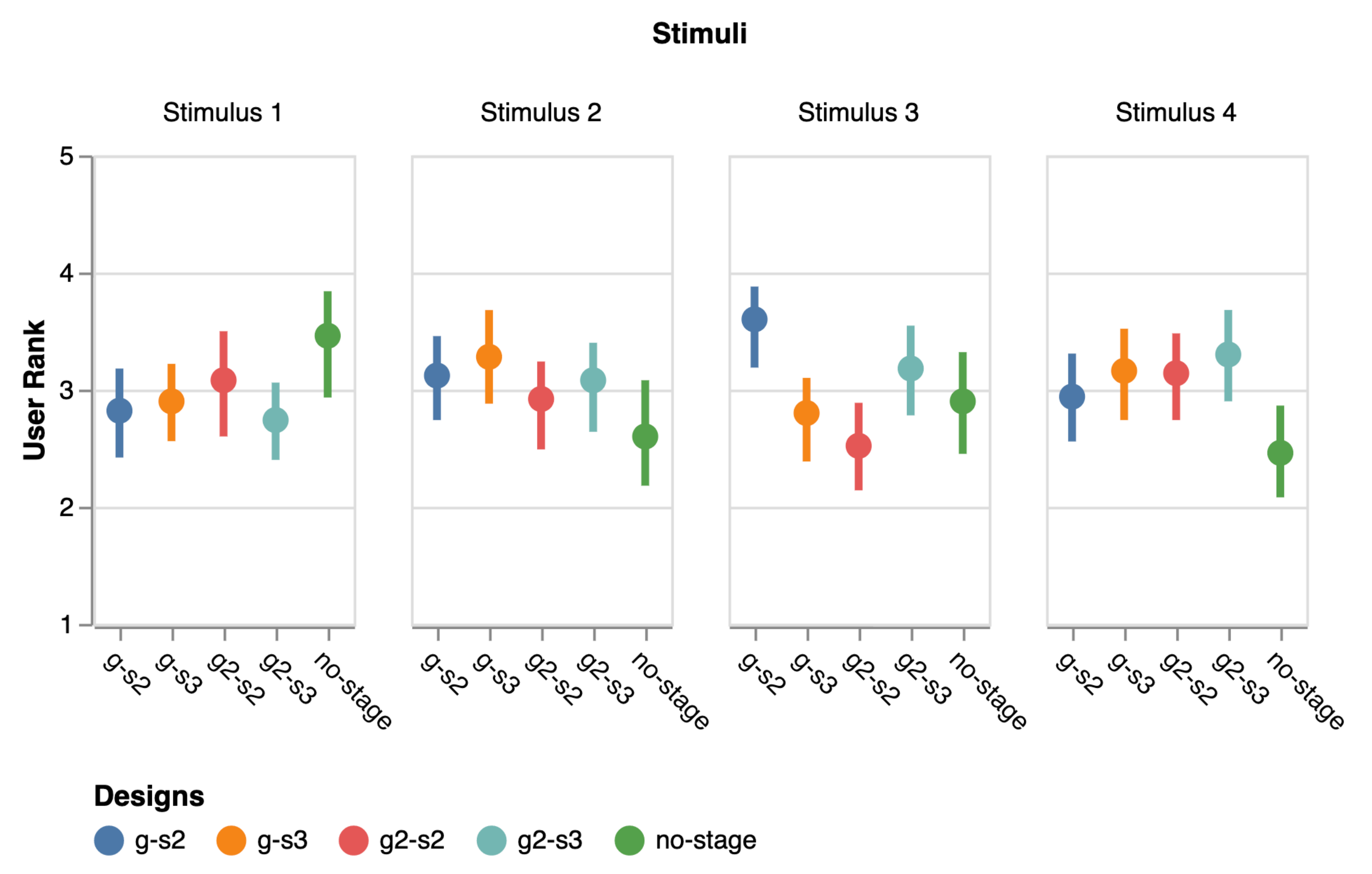}
  \vspace{-17pt}
  \caption{Bootstrapped means and 95\% confidence intervals for each animation design per stimulus.}
  \label{fig:eval}
  \vspace{-17pt}
 \end{figure}

We analyzed collected users' ranking within each stimulus instead of merging them since \gemtwo{} gave them different transition types and different animations rankings. We investigated animation designs' significant rank differences using the Friedman rank-sum test and Wilcox pairwise comparisons with Bonferroni corrections. \figref{fig:eval} shows bootstrapped means and 95\% confidence intervals of subjects' ranking across \gemtwo{}'s recommendation ranking.

\emph{Stimulus 1: Encoding \& Data Transform \& Marktype.} The first stimulus set concerns a transition that modifies a given chart's x-axis variable, aggregates its y-axis value, and changes marktype from point to bar. The main difference between Gemini's and \gemtwo{}'s staged animation designs was handling the legend: \gemtwo{} staged the legend changes according to the intermediate keyframe, but Gemini did not. On average, subjects voted \texttt{g2-s3} as the best and \texttt{no-stage} as the worst. However, we found no significant differences among the animation rankings. Multiple subjects who ranked \texttt{no-stage} fifth reported that the animation  "jumbled" the changes too quickly to understand how the data was transformed. This result suggests that \gemtwo{} can recommend staged animations for complex transitions as well as Gemini can.

\emph{Stimulus 2: Two Filters.} In this set, animations showed two filtering operations on a COVID-19 daily positive case chart; the first increased  the time range of the chart, and the second introduced data from the other regions ("Other States"). \gemtwo{}'s staged recommendations separated the data transforms into two steps: extending existing data for the new time range and introducing the other regions' data on top of the existing data. But Gemini's staged animations did not separate these data changes. Subject preferences are highest for \texttt{no-stage} but the difference is not significant difference. One possible interpretation is that these two specific filtering operations should be conveyed together since they resemble each other in terms of extending view; one extends regions, the other extends time range.

\emph{Stimulus 3: Encoding \& Data Transform.} This set involves transitions that add a y-axis for a new categorical field and aggregate data into means grouped by this field. Gemini recommended \texttt{g-s2} and \texttt{g2-s2} as the two best 2-stage animation designs. However, \gemtwo{} picked \texttt{g2-s2} as the sole best 2-stage animation due to it prioritizing keyframe sequences that apply an add encoding edit operation before an aggregate edit operation. Participants ranked \texttt{g-s2} and \texttt{g2-s2} as the worst and the best, respectively. This result suggests that \gemtwo{}'s keyframe sequence prioritization can distinguish staged animations that people may prefer.

\emph{Stimulus 4: Bin \& Aggregate.} The transitions in this set changed a 2D scatter plot to a 2D histogram by binning and aggregating. As previously mentioned, \texttt{g-s2} and \texttt{g-s3} were the best staged animation designs in \gemtwo{} as well as in Gemini. \gemtwo{} demoted \texttt{g2-s2} and \texttt{g2-s3} since they separated the bin and aggregated edit operations. The observed subjects' ranking seems to corroborate this prioritization, but none of the ranking differences are significant.

\section{Conclusion and Future Work}

In this work, we presented \gemtwo{}, a system for generating keyframe-oriented animated transitions between statistical graphics. We extended prior work on GraphScape and Gemini to make \gemtwo{} represent animations using keyframe sequences and automate animations for a given transition or keyframe sequence. \gemtwo{} is available as open-source software at \url{https://github.com/uwdata/gemini}. 

\gemtwo{} offers challenging and exciting future work opportunities. First, it is at present limited to represent animations among basic single-view Vega-Lite graphics supported by GraphScape. To support varied types of animated transitions, Gemini and GraphScape should be extended to handle more chart types. In terms of ease-of-use, one promising future direction is designing user interactions in animation authoring tools to exploit \gemtwo{}'s recommendation features. Recently, Data Animator ~\cite{data_animator} introduced novel user interfaces to author keyframe animations between data graphics. The challenge is how to seamlessly blend \gemtwo{}'s recommendation feature with such data graphic animation tools to help users explore alternative stagings while reducing their labor. Finally, additional  animation perception studies might help improve \gemtwo{}'s recommendation quality. Beyond observing users' preferences, measuring how well people understand or can identify changes through varied staged animation designs could help recommendation systems guide users toward increasingly effective designs.

\acknowledgments{
NSF award IIS-1907399 supported this work.}

\bibliographystyle{abbrv-doi}

\bibliography{gemini2}
\end{document}